\documentclass{aa}
\usepackage{epsfig}

\begin{document}

 \thesaurus{08 (02.08.1; 
 		02.13.2; 
		02.20.1; 
		09.03.1; 
		09.11.1; 
		09.19.1; 
		)}

\title{Characterizing the structure of interstellar turbulence}

\author{Mordecai-Mark~Mac Low \inst{1,3} \and Volker~Ossenkopf \inst{2}}
\authorrunning{Mac Low \and Ossenkopf}

\institute{Max-Planck-Institut f\"ur Astronomie, K\"onigstuhl 17,
  D-69117 Heidelberg, Federal Republic of Germany.
\and
1. Physikalisches Institut, Universit\"at zu K\"oln,
Z\"ulpicher Stra\ss{}e 77, D-50937 K\"oln, Federal Republic of Germany
\and
Current Address: Department of Astrophysics, American Museum of Natural 
History, 79th Street at Central Park West, New York, NY, 10024, USA}

\offprints{V.~Ossenkopf, (ossk@zeus.ph1.uni-koeln.de)}

\date{Received: ; accepted:}

\maketitle

\begin{abstract}
Modeling the structure of molecular clouds depends upon
good methods to statistically compare simulations with
observations in order to constrain the models.  Here we characterize a
suite of hydrodynamical and magnetohydrodynamical (MHD) simulations of
supersonic turbulence using an averaged wavelet transform, the
$\Delta$-variance, that has been successfully used to characterize
observations.  We find that, independent of numerical resolution and
dissipation, the only models that produce scale-free, power-law
$\Delta$-variance spectra are those with hypersonic rms Mach numbers,
above $M \sim 4$, while slower supersonic turbulence tend to show
characteristic scales and produce non-power-law spectra.  Magnetic
fields have only a minor influence on this tendency, though they tend
to reduce the scale-free nature of the turbulence, and increase the 
transfer of energy from large to small scales. The evolution of the 
characteristic length scale seen
in supersonic turbulence follows exactly the $t^{1/2}$ power-law
predicted from recent studies of the kinetic energy decay rate.

\keywords{Hydrodynamics -- Magnetohydrodynamics -- Turbulence -- ISM: clouds -- ISM: kinematics and dynamics --  ISM: structure}
\end{abstract}

\section{Introduction}

Although numerical simulations of transsonic and supersonic turbulence
appropriate to interstellar gas have been carried out for several
years now (Porter, Pouquet, \& Woodward 1992, 1994; Padoan \& Nordlund
1999; Mac Low et al. 1998; Stone, Ostriker, \& Gammie 1998) there are
only a few direct comparisons between numerical results and
astrophysical observations (e.g. Falgarone et al. 1994; Padoan et al 1999;
Rosolowsky et al. 1999).  This is mainly due to the lack of
appropriate measures applicable both to simulated and observed
structures. Measures common for turbulence studies like the power
spectrum of spatial or velocity fluctuations or the probability
distribution of velocity increments are not easily applied to
observations where their use is greatly impaired by the limitations
due to finite signal to noise ratio and limited telescope
resolution.

To obtain clues to the true physical nature of interstellar
turbulence, characteristic scales and any inherent scaling laws have
to be measured and modelled. A major problem with characterizing both
the observations and the models is to determine what scaling behaviour,
if any, is present in complex turbulent structures.  Both the velocity
and density fields need to be considered, but only the radial velocity and
column densities can be observed.

One measure useful for characterizing structure and scaling in
observed maps of molecular clouds is the $\Delta$-variance,
$\sigma^2_{\Delta}$, introduced by \cite{stutzki}. It
can better separate observational effects from the
real cloud structure than e.g. the power spectrum or fractal
dimensions. The $\Delta$-variance spectrum clearly shows
characteristic scales and scaling relations, and its logarithmic slope
can be analytically related to the spectral index of the corresponding
power spectrum.

\mbox{}\cite{stutzki} and \cite{bensch} have applied the $\Delta$-variance 
analysis to observations of the Polaris Flare and the FCRAO
survey of the outer galaxy. They found a relatively universal law
describing these clouds, with a power law structure at scales below
the cloud size and the general cloud size as the only characteristic scale
within the resolution limit.  Given the limited number of samples,
however, it is not yet possible to draw conclusions on the scaling of
turbulence in molecular clouds in general. To study the
common behaviour and differences between several clouds and
interstellar regions the analysis of more and larger maps obtained
with a good signal-to-noise ratio is required.

In order to understand the physical significance of the
characterization of the observational maps by $\Delta$-variance
spectra, we apply here the same analysis to simulated gas
distributions resulting from MHD models.  In this first paper, we try
to get a general feeling for the scaling behaviour in
different models, and for the influence of the different parameters and
numerical approaches on the produced structures. We only perform a
qualitative comparison to the observations here.  In a subsequent
paper we will attempt to make a detailed fit of
several observed regions using MHD models including the solution
of the radiative transfer problem.

\section{Structure measure by the $\Delta$-variance}

\subsection{Definitions}
 
The $\Delta$-variance was comprehensively introduced by \cite{stutzki}.
We will repeat here only the formalism essential for the further
analysis in this paper.

The $\Delta$-variance is a type of averaged wavelet transform that
measures the variance in an $E$-dimensional structure $f(\vec{r})$
filtered by a spherically symmetric down-up-down function of varying
size (Zielinsky \& Stutzki 1999). It is defined by
\begin{equation}
\sigma_\Delta^2(l)= \int_{-\infty}^\infty \left( (f(\vec{r})-\langle f
\rangle) * {\bigodot}_l(\vec{r}) \right)^2 d\vec{r} 
\end{equation}
where, the $*$ stands for a convolution and $\bigodot_l$ describes the 
down-up-down function with the length $l$ of each step
\begin{equation}
{\bigodot}_l(\vec{r})= {\cal V}_E^{-1} ({2 \over l})^E \left\{ \begin{array}{ll}
1 & |\vec{r}| \le l/2\\
-1/(3^E-1) & l/2 < |\vec{r}| < 3l/2\\
0 & |\vec{r}| > 3l/2 \end{array} \right.
\end{equation}
with ${\cal V}_E$ being the volume of the $E$-dimensional unit sphere.

Thus, the $\Delta$-variance measures the amount of structural variation
on a certain scale, e.g. in a map or three-dimensional distribution. A familiar,
slightly different kind of variance defined for one-dimensional problems is
the Allan-variance commonly used for stability investigation (\cite{schieder}).
In contrast to the $\Delta$-variance, it works with a non-symmetric up-down 
filter.

Instead of convolving the structure in ordinary space with a filter function
one can carry out the $\Delta$-variance analysis in Fourier space
by simple multiplication. This directly relates the $\Delta$-variance
to the power spectrum of a structure. If $P(k)$ is the radially averaged
power spectrum of the structure $f(\vec{r})$, the $\Delta$-variance is
given by
\begin{equation}
\sigma_\Delta^2(l)= \int_0^\infty P(k)\; | \tilde{\bigodot}_l(k) |^2 k^{E-1} dk
\end{equation}
where $\tilde{\bigodot}_l$ is the Fourier transform of the $E$-dimensional 
down-up-down function with the scale length $l$, and we are using $k$
to denote the spatial frequency or wavenumber.

If the power spectrum is given by a simple power law, $P(k)\propto
k^{-\zeta}$, the $\Delta$-variance also follows a power law
$\sigma_\Delta^2 \propto l^\alpha$ with $\alpha={\zeta-E}$ within 
the exponential range
$0\le \zeta < E+4$.  (To avoid confusion with the
ratio between thermal and magnetic pressure we use $\zeta$ for the
power spectral index rather than the variable $\beta$ used by
\cite{stutzki} and \cite{bensch}.)  The main advantages of the
$\Delta$-variance compared to the direct computation of the Fourier
power spectrum are the clear spatial separation of different effects
influencing observed structures like noise or finite observational
resolution, and the robustness against singular variations due to
the regular filter function.

Furthermore, for most astrophysical structures, where a periodic
continuation is not possible, \cite{bensch} has shown that the
periodicity artificially introduced by the Fourier transform can lead
to considerable errors. Here, even the $\Delta$-variance has to be
determined in ordinary space. For the simulations examined
in this paper, periodic wrap around is not a problem because it is
already explicitly assumed, so we apply the faster Fourier method
to determine their $\Delta$-variance spectra.

\begin{figure}
\epsfig{file=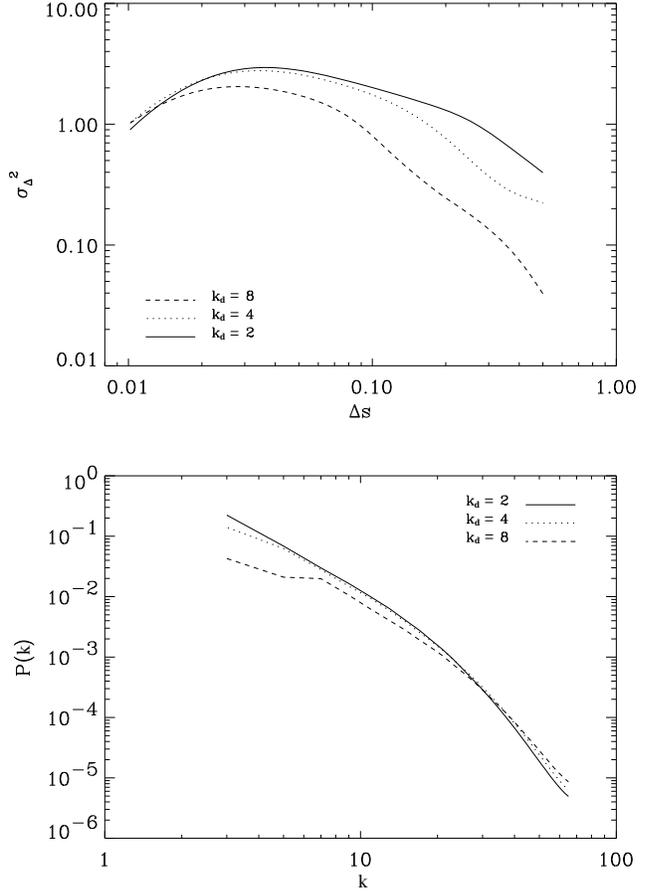,angle=0,width=\columnwidth}
\caption{Comparison of 3D $\Delta$-variance spectra to Fourier power
spectra for $128^3$ models of turbulence driven at wavenumbers of $k_d =
2$ (solid), $k_d = 4$ (dotted), and $k_d = 8$ (dashed), demonstrating 
that the analytic relation between power spectrum slope and $\Delta$-variance
spectrum slope also holds for the local behavior of structures
not showing a straight power-law.}
\label{powdelta}
\end{figure}
In Fig. \ref{powdelta} we compare the power spectrum and $\Delta$-variance
for three simulations described below. They have different driving
scales, and therefore each shows a different characteristic scale
visible as a turn-over at large lags in the $\Delta$-variances and
at small wavenumbers in the power spectrum, respectively. At smaller
lags and higher wavenumbers power laws can be seen in both cases (with their
slopes related by the analytic relation given above). A steep drop-off
follows at the smallest scales indicating the resolution limit of the 
simulation. The power laws are equivalent in both cases, but the
characteristic scale at one end and the resolution limit at the other
end of the spectrum can be more clearly seen in the $\Delta$-variance.
The smooth spatial filter function in the $\Delta$-variance analysis still
provides a good measure for the behavior at large scales whereas the
power spectrum suffers from the low significance of the few remaining points
there. 

The $\Delta$-variance analysis of astronomical maps was extensively 
discussed and demonstrated by \cite{bensch}. In all the observations
analyzed by them, the total cloud size was the 
only characteristic scale detected by means of the $\Delta$-variance.
Below that size they found a self-similar scaling behaviour reflected
by a power law with index $\alpha=0.5 \dots 1.3$ corresponding
to a Fourier power spectral index $\zeta=2.5 \dots 3.3$. The analysis of 
further maps will be discussed in future work.

\subsection{Two-dimensional maps and three-dimensional structures}

In the application to molecular cloud structure simulations we
have to restrict the analysis either to the three-dimensional
structure or to the two-dimensional projection of the
structure which would be astronomically observed, e.g. in optically thin
lines or the FIR dust emission.

\mbox{}\cite{stutzki} have shown that the spectral index of the power spectrum
$\zeta$ for an spatially isotropic structure remains constant on
projection. This means that the projected map of three-dimensional density 
structures shows the same $\zeta$ as the original structure as
long as we assume that the astronomical structure is on the
average isotropic. Consequently the slope of the $\Delta$-variance
grows by 1 in projection.

\begin{figure}
\epsfig{file=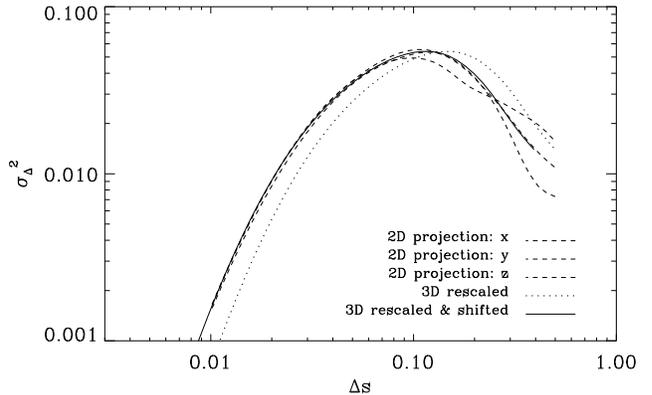,angle=0,width=\columnwidth}
\caption{Demonstration of the equivalence of 2D projected and properly
transformed 3D $\Delta$-variance spectra.  Projected
$\Delta$-variances spectrum of a model density cube (model C from Mac
Low et al. 1998 at $t = 0.1 t_s$, where $t_s = L/c_s$ is the sound-crossing time) are shown. The dotted line is the
corresponding three-dimensional $\Delta$-variance multiplied by the
local lag. The solid line is additionally shifted by a factor $\pi/4$
to correct for the average length reduction on projection.
}
\label{two_d-three_d}
\end{figure}
In Fig. \ref{two_d-three_d} we demonstrate this for a simulation where 
we determined the $\Delta$-variance of the three-dimensional structure 
and the $\Delta$-variances of the three perpendicular projections. 
The dashed lines show the three projected $\Delta$-variances. Now
we multiply the three-dimensional $\Delta$-variance by the 
abscissa values to obtain the same local slope as measured in
two dimensions (dotted line). 
However, we still have to correct for the scale length of the measure.
The length of an arbitrary three-dimensional vector is reduced on
projection to two dimensions by a factor $\pi/4$ on the average. Therefore,
we adjust the local scale by this factor for the $\Delta$-variance
determined in three dimensional space. The resulting plot
is shown as the thick solid line in Fig. \ref{two_d-three_d}.
We obtain exactly the same general behaviour
as for the projected maps. The equivalent plot for numerous other models
verified this as a general behavior. Hence, we can either consider the
three-dimensional variance or the projected variances and can
simply translate them into each other.

The treatment of the three-dimensional variances is favourable from the
viewpoint that it measures exactly the scales as they occur in
the density structure. However, the projected maps are favourable to 
have a means of direct comparison to astronomical observations.
Putting the relation to the observations at first priority, we
will show in the following the variances translated to the
two-dimensional behaviour and we will only mention the physical
three-dimensional scales if they appear to be especially prominent.

In this paper, we will restrict ourselves to simple projections 
taking them as representations of the integrated map of optically 
thin lines or optically thin continuum emission.
We will not treat the full radiative transfer problem which had to
be solved for a general treatment. Optical depths effects in
fractal and random structures were discussed by
\cite{ossk-puebla} and  they will be taken into account in a subsequent
paper dealing with the simulation of certain molecular clouds.

\begin{figure}
\epsfig{file=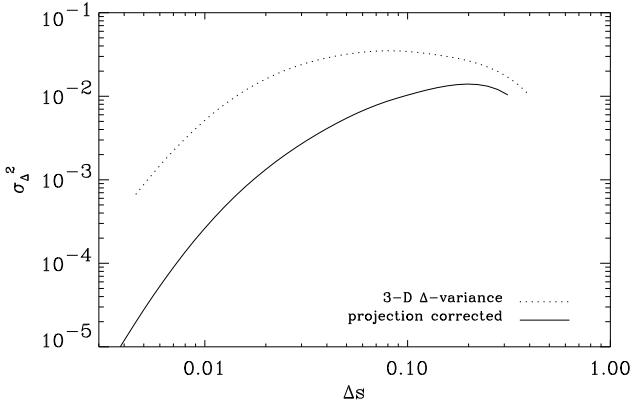,angle=0,width=\columnwidth}
\caption{$\Delta$-variance of a decaying hydrodynamic model 
determined in 3-D and rescaled for projection effects as if
it were measured as a 2-D map.}
\label{2d_correction}
\end{figure}

As a side-result of this comparison we find however, that the 
treatment of maps instead of three-dimensional cubes by observers 
can easily lead to a misinterpretation of the structure scaling.
Fig. \ref{2d_correction} compares the three-dimensional
$\Delta$-variances computed in 3-D and the same variance corrected 
for 2-D projection as it could be measured by an observer
for a hydrodynamic decaying turbulence model. Whereas the
plot for 3-D only shows a broad distribution of structures,
the human eye tries to see in the 2-D curve at least a reasonable range with
a power law between 0.03 and 0.2. Except for the smallest lags
dominated by numeric viscosity as discussed below the plot
is quite similar to variances obtained e.g. by \cite{stutzki}
for molecular clouds. Hence, noncritical observers might be forced to see
a self-similar behaviour even if there is no strong indication
for a power law.

\section{Numerics}

\subsection{Computations}

We use simulations of uniform decaying and driven turbulence with and
without magnetic fields described by Mac Low et al.\ (1998) in the
decaying case and by Mac Low (1998) in the driven case.  These
simulations were performed with the astrophysical MHD code
ZEUS-3D\footnote{Available by registration with the Laboratory for
Computational Astrophysics of the National Center for Supercomputing
Applications at the email address lca@ncsa.uiuc.edu} (Clarke 1994).
This is a three-dimensional version of the code described by Stone \&
Norman (1992a, b) using second-order advection (Van Leer 1977), that
evolves magnetic fields using constrained transport (Evans \& Hawley
1988), modified by upwinding along shear Alfv\'en characteristics
(Hawley \& Stone 1995).  The code uses a von Neumann artificial
viscosity to spread shocks out to thicknesses of three or four zones
in order to prevent numerical instability, but contains no other
explicit dissipation or resistivity.  Structures with sizes close to
the grid resolution are subject to the usual numerical dissipation,
however.

In this paper, we attempt to use these simulations to derive some of
the observable properties of supersonic turbulence.  Although our
dissipation is clearly greater than the physical value, we can still
derive useful results for structure in the flow that does not depend
strongly on the details of the behavior at the dissipation scale.
Such structure exists in incompressible hydrodynamic turbulence
(e.g. Lesieur 1997).  In Mac Low et al.\ (1998) it was shown that the
energy decay rate of decaying supersonic hydrodynamic and MHD
turbulence was independent of resolution with a resolution study on
grids ranging from $32^3$ to $256^3$ zones.  Because both numerical
dissipation and artificial viscosity act across a fixed number of
zones, increasing resolution yields decreasing dissipation.  The
results we describe in this paper suggest that in some cases
observable features may be independent enough of resolution, and thus
of the strength of dissipation. Despite the limitations of our method 
we can therefore draw quantitative conclusions.  Again, we support this
assertion by appealing to resolution studies whenever possible.

The simulations used here were performed on a three-dimensional,
uniform, Cartesian grid with side $L = 2$, extending from -1 to 1 with
periodic boundary conditions in every direction. For convenience, we
have normalized the size of the cube to unity in the analyses
described here, so that all length scales are in fractions of the cube
size.  An isothermal equation of state was used in the computations,
with sound speed chosen to be $c_s = 0.1$ in arbitrary units.  The
initial density and, in relevant cases, magnetic field were both
initialized uniformly on the grid, with the initial density $\rho_0 =
1$ and the initial field parallel to the $z$-axis.

The turbulent flow is initialized with velocity perturbations drawn
from a Gaussian random field determined by its power distribution in
Fourier space, following the usual procedure. As discussed in detail
in Mac Low et al.\ (1998), it is reasonable to initialize the decaying
turbulence runs with a flat spectrum with power from $k_d = 1$ to $k_d =
8$ because that will decay quickly to a turbulent state.  Note that
the dimensionless wavenumber $k_d = L/\lambda_d$ counts the number of
driving wavelengths $\lambda_d$ in the box.  A fixed pattern of
Gaussian fluctuations drawn from a field with power only in a narrow
band of wavenumbers around some value $k_d$ offers a very simple
approximation to driving by mechanisms that act on that scale. To
drive the turbulence, this fixed pattern was normalized to produce a
set of perturbations $\delta\vec{\nu}(x,y,z)$, and at every time step
add a velocity field $\delta\vec{v}(x,y,z) = A \delta\vec{\nu}$ to the
velocity $\vec{v}$, with the amplitude $A$ now chosen to maintain
constant kinetic energy input rate, as described by Mac Low (1998). 

\subsection{Resolution Studies}

\begin{figure}
\epsfig{file=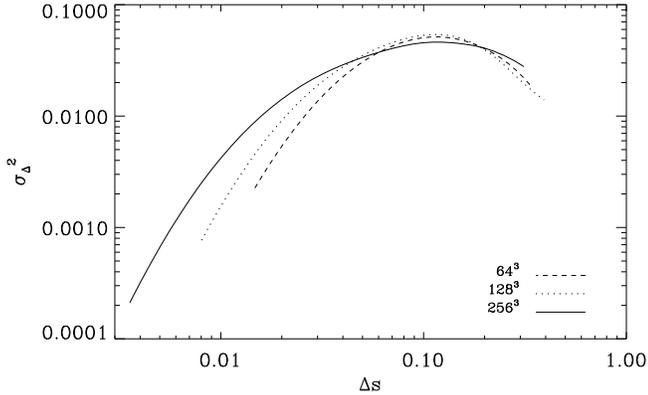,angle=0,width=\columnwidth}
\caption{$\Delta$-variance of the density cubes (translated to
two-dimensions) for a model of decaying hydrodynamic 
turbulence at a time $t/t_s = 0.1$, computed using numerical resolutions of
$64^3$, $128^3$, and $256^3$.
}
\label{numresolution}
\end{figure}

In Figure~\ref{numresolution} we show how numerical resolution, or
equivalently the scale of dissipation, influences the
$\Delta$-variance spectrum that we find from our simulations.  We test
the influence of the numerical resolution on the structure by
comparing a simple hydrodynamic problem of decaying turbulence
computed at resolutions from $64^3$ to $256^3$, with an initial rms
Mach number $M = 5$ (Model D from Mac Low et al. 1998).

In contrast to the results from \cite{maclow} which showed little
dependence of the energy dissipation rate on the
numerical resolution, we find here remarkable differences
in the scaling behaviour of the turbulent structures.
At small scales we find a very similar decay in the
relative structure variations up to scales of about 10
times the pixel size (0.03, 0.06, and 0.1 for the
resolutions $256^3$, $128^3$, and $64^3$, respectively) in 
all three models. This constant length range starting from
the pixel scale clearly identifies this decay as an artifact
from the simulations which can be attributed to the numerical
viscosity acting at the smallest available size scale.

Another very similar behaviour can be observed at the
largest lags where the relative structure variations decay
for all three simulations on a length scale covering a factor
two below half the cube size. This structure reflects the
original driving of the turbulence with a maximum wavenumber
$k_d=8$ that manifests itself in the production of structure
on the corresponding length scale. Only for the $256^3$ cubes
we find a range of an approximately self-similar behaviour
at intermediate scales that is not yet smoothed out by the
influence of numerical viscosity. 

Structures larger than at most half the cube size are 
suppressed by the use of periodicity in the
simulations. Together with the viscosity range of about 10
pixels there is only a scale factor about 10, 5 or 3
remaining for the three different resolutions where
we can study true structure not influenced by the limiting
conditions of the numerical treatment. For the derivation
of reliable scaling laws, we must therefore use at least
simulations on the $256^3$ grid. On the other hand
we know, however, that the limits of the observations
also constrain the scaling factor for structure
investigations in observed maps to at most a factor 10 
in general (\cite{bensch}).

Although we have plotted here only the results for a
hydrodynamic model there are no essential differences
to the resolution dependence when magnetic fields are included
as discussed below.

\subsection{Statistical Variations}
Another question concerns the statistical significance of the 
structure in the simulations. Since each simulation and even
each time step provides another structure there is a priori
no reason to believe that a statistical measure like the
$\Delta$-variance is about the same for each realization
of a given HD/MHD problem.

Restricted by the huge demand for computing power in each
simulation we cannot provide a statistically significant analysis of 
many realizations for each problem. However, we will try to
provide some general clues for the uncertainty of the
$\Delta$-variance measured for a certain structure.

A first impression can be obtained from the differences in the
three projections of one cube in Fig. \ref{two_d-three_d}. Because 
each projection provides an independent view on the three-dimensional
structure their variation can be considered a rough measure
for the statistical significance of the $\Delta$-variance plots.
We see that the curves agree well up to lags of about a quarter
of the cube size but deviate considerably at larger lags.
This is explained by the number of structures contributing to
the variations at each scale. Whereas we find many small fluctuations 
dominating the variance at small scales there is in general only
one main structure responsible for the variance at the largest scale.
Its different appearance from different directions then produces
the uncertainty in the $\Delta$-variance there.

Looking at the variance determined in three dimensions in Fig. 
\ref{two_d-three_d} we see however that it provides already a kind of 
average over the three projected functions. Analyzing the three-dimensional
cubes thus removes already part of the statistical variations
that could be seen by an observer when looking at the two-dimensional
projections only. The statistical
uncertainty is reduced for the $\Delta$-variances determined
in three dimensions considered below. 

As another estimate for 
the uncertainty in this case we study the variances for
different time steps in the evolution of a continuously driven
hydrodynamic model. In the evolution of the simulation 
different structures are produced which should behave
statistically equal since the general process of their formation
and destruction remains the same.

\begin{figure}
\epsfig{file=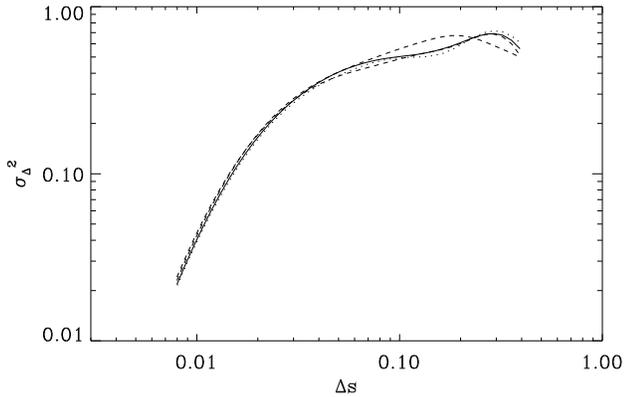,angle=0,width=\columnwidth}
\caption{
Different time steps in the evolution of a driven model (model HE2
from Mac Low 1999 with $k_d=2$ driving) which should remain
statistically stationary, showing that the $\Delta$-variation spectra
don't vary much due to turbulent evolution. The first time is $0.4
t_s$, and succeeding plots are separated by times of $ 0.05 t_s$, or
0.75 the crossing time at the rms velocity.
}
\label{driven_timesteps}
\end{figure}

Fig. \ref{driven_timesteps} shows four different timesteps in an HD
model driven at wavenumber $k_d=2$ each separated by 0.75 the box
crossing time at the rms velocity. The variations even at larger
scales are much less than in Fig. \ref{two_d-three_d}.  It appears
that the $\Delta$-variance does a good job of characterizing invariant
properties of the structure.  Only for high accuracy determinations of
the slope or the reliable identification of self-similarity ensemble
averages should be taken by computing many realizations.

\section{Results}

\subsection{Decaying hydrodynamic turbulence}

We have computed the $\Delta$-variance spectra for two models of
decaying hydrodynamic turbulence, one with initial rms Mach number $M
= 5$, noted as Model~D in Mac Low et al. (1998), and one otherwise
identical model with initial $M = 50$, not published before.  As noted
above, these models were excited with a flat-spectrum pattern of
velocity perturbations, which would correspond to a rather steep
spectrum $\sigma_{\Delta}^2 \propto L^{-2}$ in 2D or $\propto L^{-3}$ in 3D,
respectively.  Both were run at a resolution of $256^3$.
\begin{figure}
\epsfig{file=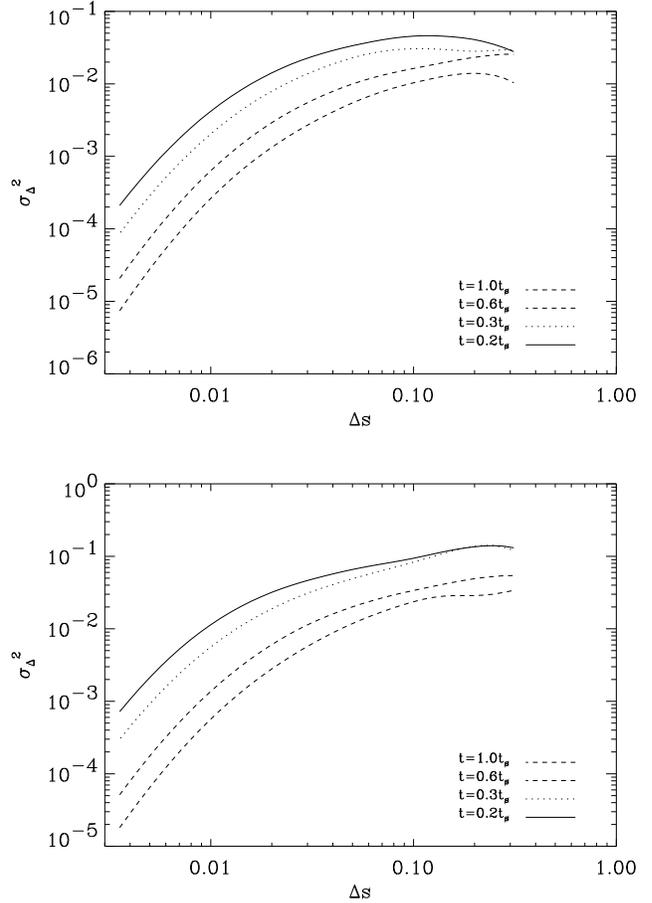,angle=0,width=\columnwidth}
\caption{Time sequence of decaying turbulence originally driven by M=5
(upper plot, model D from Mac Low et al. 1998), and M=50 (lower plot), 
at times in units of the sound crossing time $t_s$.
}
\label{decay_fig}
\end{figure}

The first time steps in Figure~\ref{decay_fig} show that only hypersonic 
turbulence provides a self-similar behavior, indicated by a power-law
$\Delta$-variance spectrum.  In this case, there appears to be
structure corresponding to a power-law spectrum of $k^{-2.5}$,
somewhat steeper than the $k^{-2}$ that would be expected from a
simple box full of step-function shocks, but approaching the steepness
observed for real interstellar clouds.  When the turbulence decays
to supersonic rms velocities at later times, or in the model having
only supersonic initial velocities, the spectrum indicates no self-similar 
structure but a distinctive physical scale that evolves with time to larger
sizes.

We speculate that a physical explanation for this observation might be
drawn from the nature of dissipation in supersonic turbulence.  Energy
does not cascade from scale to scale in a smooth flow through
wavenumber space as is assumed by analyses following Kolmogorov (1941)
for subsonic turbulence.  Rather, energy on large scales is directly
transferred to scales of the shock thickness by shock fronts, and
there dissipated.  As a result, energy is not added to small and
intermediate scale structures at the same rate that it is dissipated.
Combined with a fairly steep power spectrum, this means that the
smaller scale structures will be lost to viscous dissipation first,
moving the typical size to larger and larger scales.

We can quantify the change in typical scale by simply fitting a
power-law to the lag $L_{pk}$ at which $\sigma_{\Delta}^2$ reaches a peak.
This was done for the 3D $\Delta$-variance where the peak appears
more prominent than in Fig. 6 and represents the true length scale 
without projection effects. For the model starting at Mach 5 we find a 
variation in time of $L_{pk} \propto t^{q}$ with $q = 0.51$.  

These decaying turbulence models were found by Mac Low et al. (1998)
to lose kinetic energy at a rate $E_{\rm kin} \propto t^{-\eta}$, with
$\eta \simeq 1$.  However, Mac Low (1999) showed that driven
hydrodynamic turbulence dissipates energy $E_{\rm kin} \propto
v^3/\ell$, corresponding to a kinetic energy decay rate of $\eta = 2$
if the effective decay length scale $\ell$ were independent of time. From 
this observation, a time dependence of $\ell \propto t^{1/2}$ was
deduced.  Mac Low (1999) also showed that the characteristic driving
length-scale $1/k_d$ was the most likely identification for $\ell$.
Identifying $\ell$ for decaying turbulence with the length scale
containing the most power in the $\Delta$-variance spectrum $L_{pk}$
seems natural, and yields excellent agreement in the time-dependent
behavior of the length scale, since $L_{pk} \propto t^{0.51}$.

\label{decay_sect}

\subsection{Driven hydrodynamic turbulence}

\begin{figure}
\epsfig{file=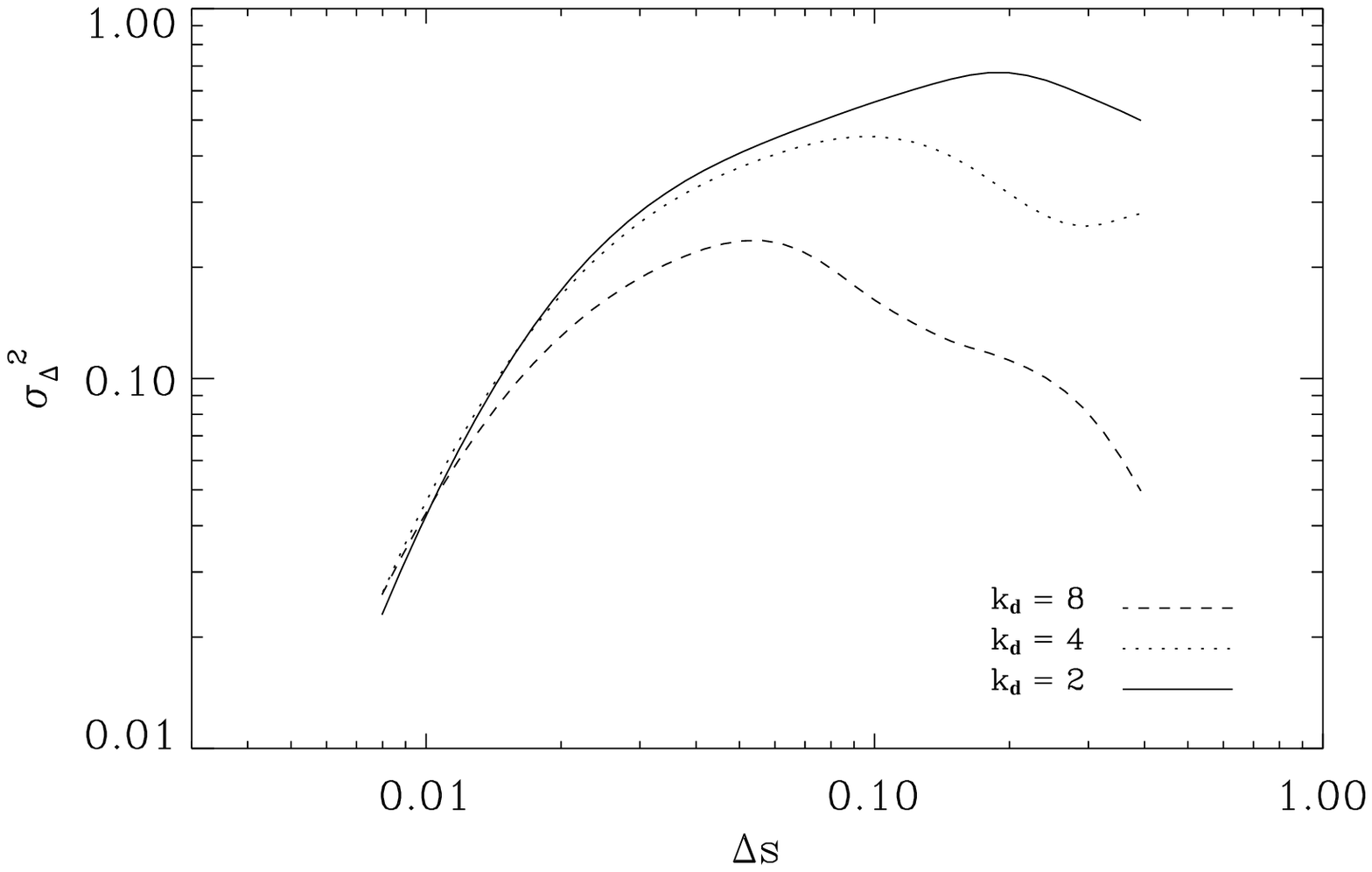,angle=0,width=\columnwidth}
\epsfig{file=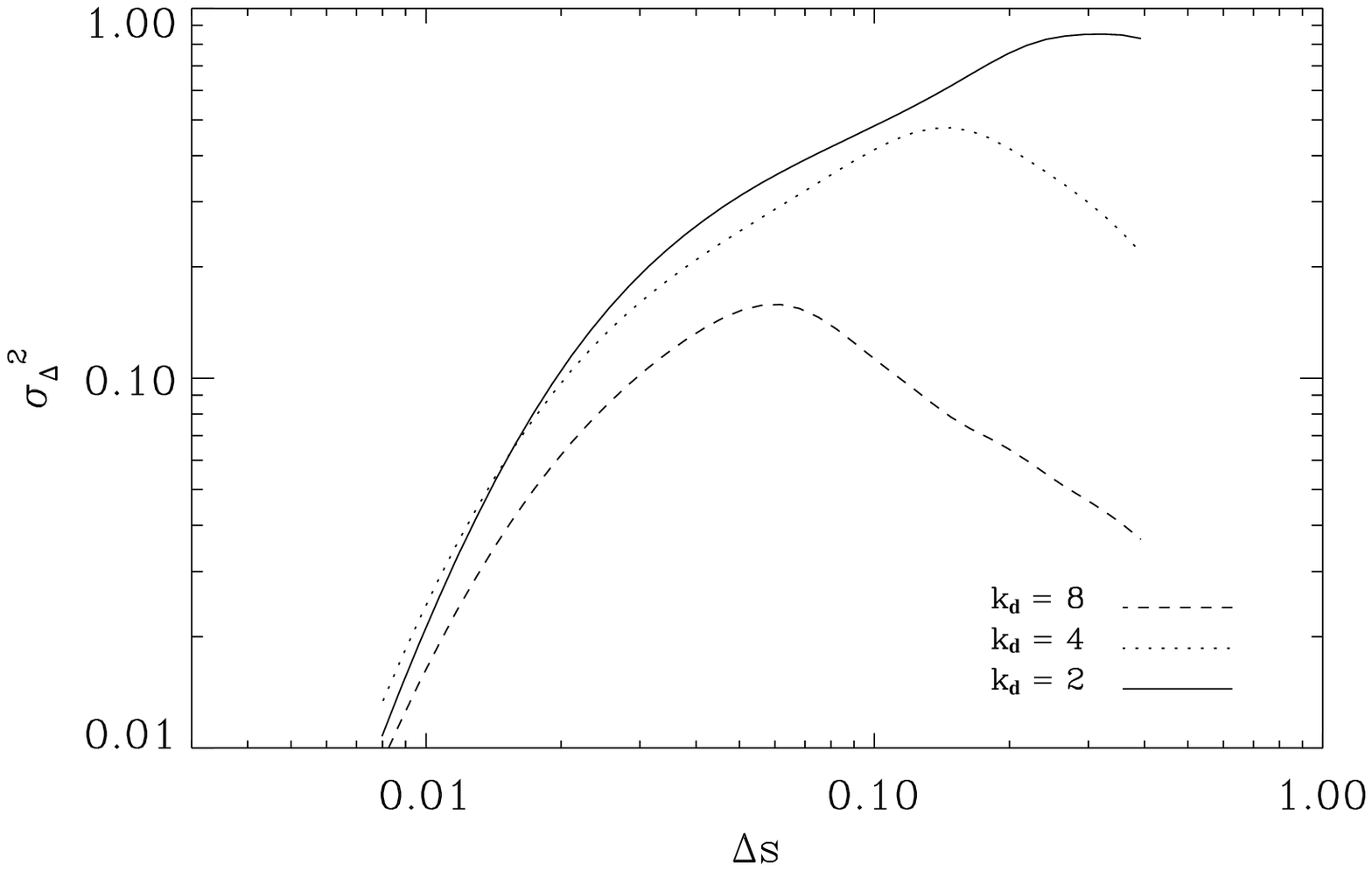,angle=0,width=\columnwidth}
\caption{$\Delta$-variance spectra of hydrodynamical models
continuously driven at $k_d = 2$, 4, and 8. In the upper part,
the turbulence is driven at hypersonic velocities
(models HE2, HE4, and HE8 from Mac Low 1999); in the lower part the 
driving energy is reduced by a factor 10 (models HC2, HC4, and HC8).
}
\label{driven_fig}
\end{figure}

In Figure~\ref{driven_fig} we show the $\Delta$-variance spectra for
models of supersonic hydrodynamic turbulence driven with a fixed
pattern of Gaussian random perturbations having only a narrow range of
wavelengths and two different energy input rates. The driving wavelengths 
are 1/2, 1/4, and 1/8 of the cube size, corresponding to driving 
wavenumbers of $k_d = 2$, 4, and 8. In the upper graph (models HE2,
HE4, and HE8 from Mac Low 1999), the driving power is by
a factor 10 higher than in the lower graph (models HC2, HC4, and HC8).
The equilibrium rms Mach numbers here are 15, 12, and 8.7,
for the high energy models driven with $k_d = 2$, 4, and 8 respectively,
and 7.4, 5.3, and 4.1 for the low energy simulations.
All of these models were run at $128^3$ resolution.

All spectra show a prominent peak characterizing the dominant
structure length. It is obviously related to the scale on which the
turbulence is driven but the exact position depends on the energy
input rate. Whereas all peak positions in the strongly
driven case are at about 0.5 times the driving wavelength
(correcting the scales from Fig. \ref{driven_fig} by the projection
factor $4/\pi$), they change in the lower graph from 0.8 times
the driving wavelength for $k_d=2$ to 0.6 $\lambda_d$ for $k_d=8$.
Thus, only the strongly hypersonic models provide a constant relation
between the driving scale and the dominant scale of the density
structure. 

Below the peak scale, a power-law distribution of structure is observed,
while above this scale, the spectrum drops off very quickly.  The
power-law section of the spectrum has a slope between 0.45 for the
high Mach number models and 0.75 for the lower Mach numbers, corresponding to a 
power spectrum power law of $k^{-2.45}\dots k^{-2.75}$. This agrees with
the slope observed in the case of hypersonic decaying turbulence  and is 
well in the range observed in real molecular clouds. 
 
Further simulations should systematically study the transition from 
supersonic to hypersonic velocities in driven models to find the
critical parameters for the onset of a self-similar behaviour and
the exact relation between the peak position, the driving scale,
and the viscous dissipation length in this case.
\label{driven_sect}

\subsection{MHD models}

\begin{figure}
\epsfig{file=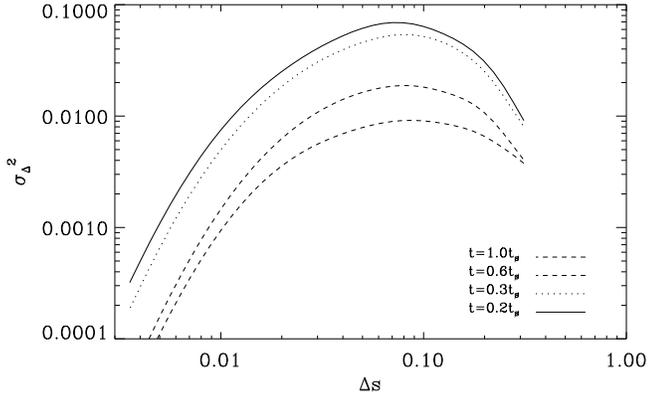,angle=0,width=\columnwidth}
\caption{Decaying turbulence in an MHD model (model Q in Mac Low et
al. 1998) with strong magnetic field at times in units of the sound
crossing time $t_s$. 
\label{mhd_decay}}
\end{figure}

Now we can examine what happens when magnetic fields are introduced to
models of both decaying and driven turbulence.  In Figure~\ref{mhd_decay} 
we begin by examining the $\Delta$-variance spectra of a decaying
model with $M=5$ and initial rms Alfv\'en number $A=1$, equivalent to
a ratio of thermal to magnetic pressure $\beta = 0.08$.  This $256^3$
model was described as Model~Q in Mac Low et al. (1998).  

No power law behavior is observed, with the spectra showing a
uniformly curved shape remarkably devoid of distinguishing features.
We emphasize that this behavior is preserved through a resolution
study encompassing a factor of four in linear resolution, suggesting
that it is not simply due to numerical diffusivity, but rather is a
good characterization of the structure of a strongly magnetized
plasma.  Thus we conclude that self-similar, power-law behavior is
{\em not} a universal feature of MHD turbulence, and that observations
showing such curved $\Delta$-variance spectra may reflect the true
underlying structure, rather than being imperfect observations of
self-similar structure.
The magnetic field tends to transfer power from larger to smaller
scales quickly, overpowering the evolution of the characteristic
driving scale seen in the hydrodynamical models. 

\begin{figure}
\epsfig{file=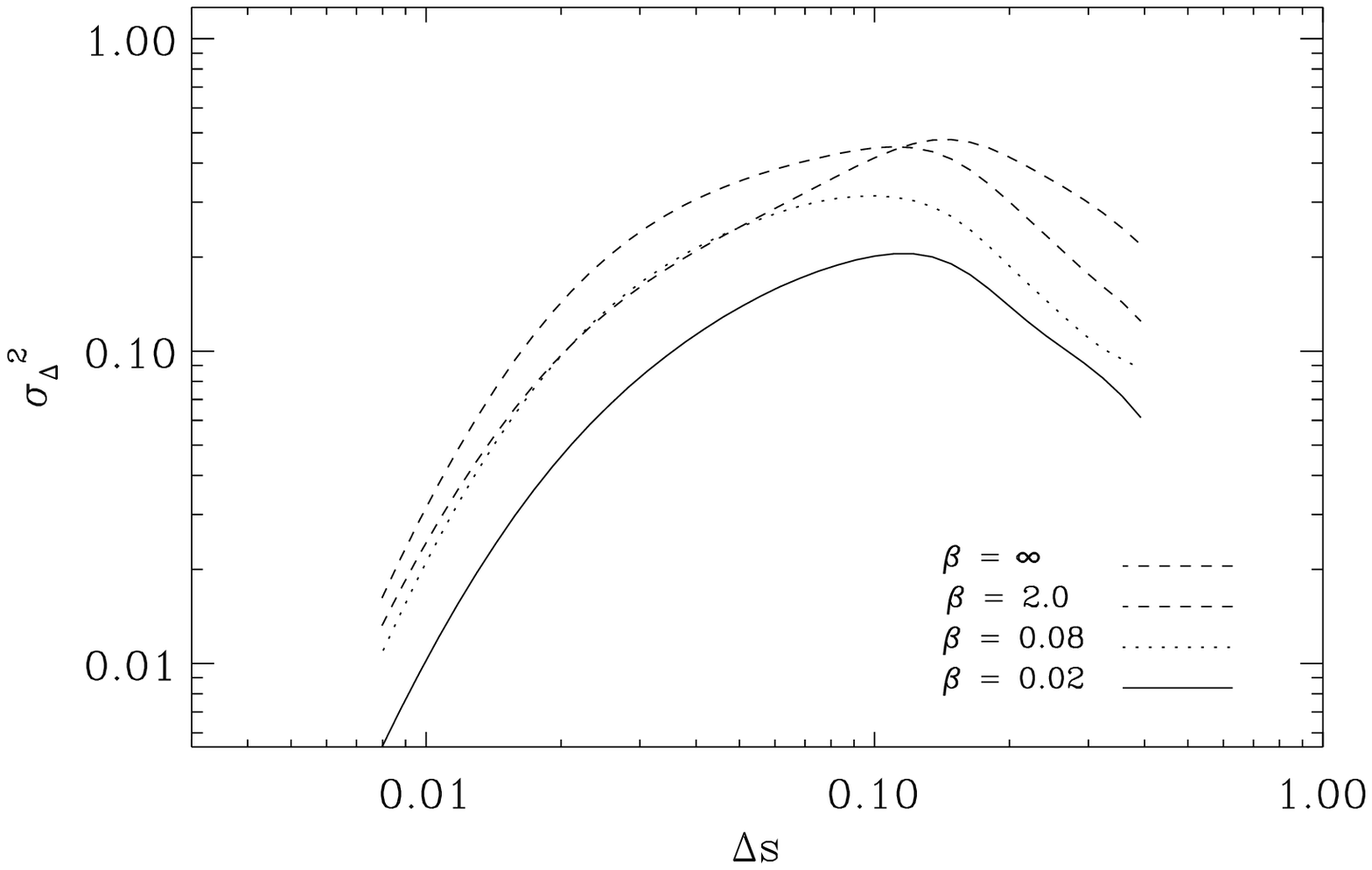,angle=0,width=\columnwidth}
\epsfig{file=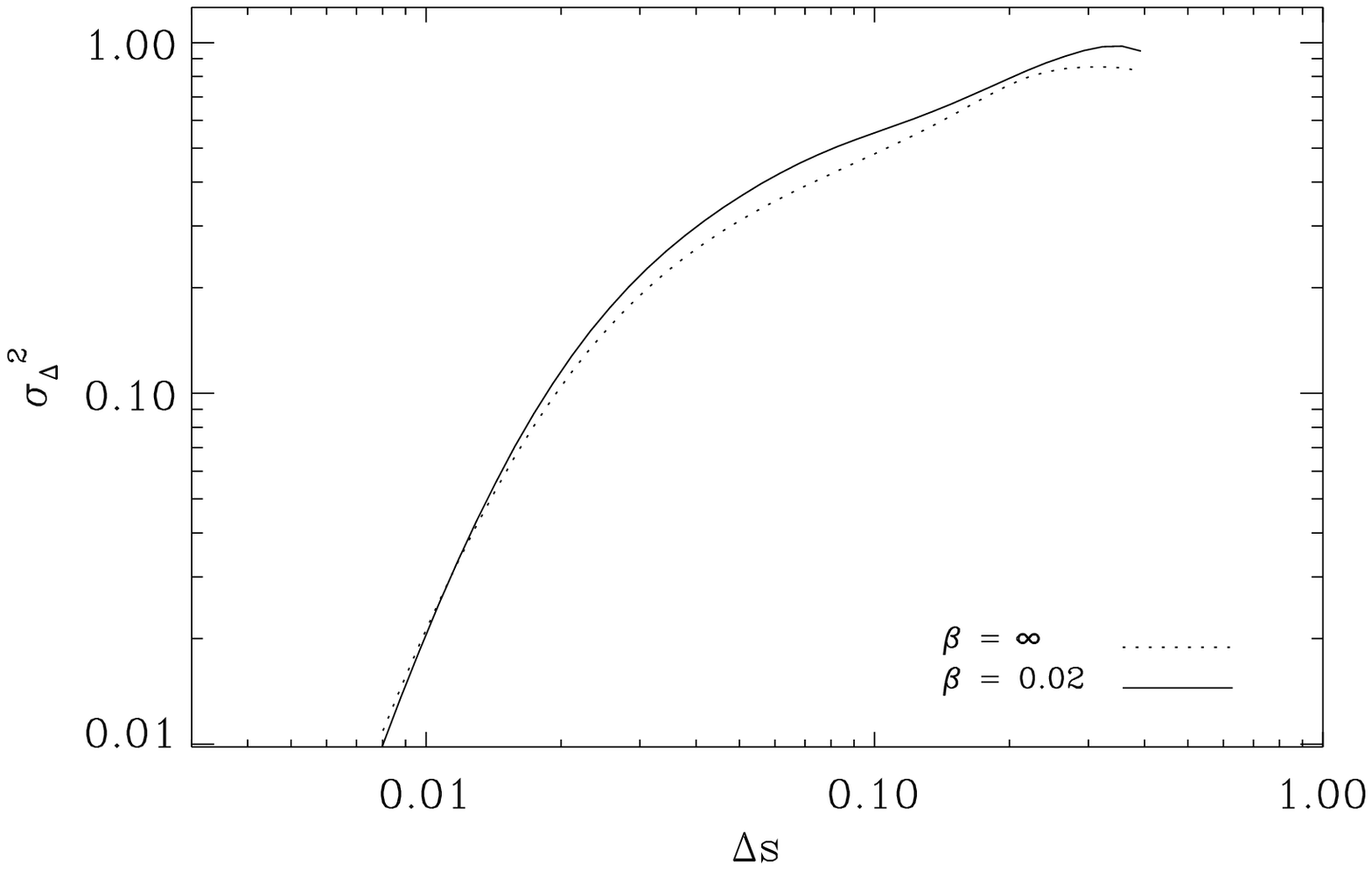,angle=0,width=\columnwidth}
\caption{ Influence of the magnetic field strength on the structure
produced in MHD driven turbulence models with (above) $k_d = 4$ driving
and (below) $k_d=2$ driving.  Note that $\beta = \infty$ is a
hydrodynamical model.  The magnetic fields tend to transfer energy
from larger to smaller scales, though the effects are not huge.
\label{mhd_driven}}
\end{figure}

A similar behavior is  visible in the driven turbulence models 
shown in Fig. \ref{mhd_driven}.
In the upper part of the figure the $\Delta$-variance 
spectra for three $128^3$ models with driving wavenumber $k_d=4$ and 
ratios of thermal to
magnetic pressure of $\beta = 0.02$, 0.08, and 2.0 (models MC4X,
MC45, and MC41 as described by Mac Low 1999) are shown along with a
hydrodynamical model ($\beta = \infty$) with identical driving (HC4).
The MHD models all have equilibrium rms Mach number $M \sim 5$; 
their equilibrium rms Alfv\'en
numbers are about 0.8, 1.6, and 8 respectively. In the lower graph
we have plotted the equivalent extreme cases of $\beta=0.02$ and 
$\beta=\infty$ for the $k_d=2$ driving.

We find again that the magnetic
fields have some tendency to transfer energy from large to small
scales, presumably through the interactions of non-linear MHD waves.
The more energy that is transferred down to the dissipation scale, the
less power is seen in the $\Delta$-variance spectra, suggesting that
the strong field ($\beta = 0.02$) is more efficient at energy transfer
than the weaker, higher $\beta$ fields.  The larger-scale $k_d=2$
driving admittedly shows much less drastic effects than the $k_d=4$
driving, emphasizing that the magnetic effects are secondary in
comparison to the nature of the driving.

This transfer of energy to smaller scales has implications for the
support of molecular clouds.  There have been suggestions by
Bonazzola et al. (1987) and L\'eorat et al. (1990) that turbulence can
only support regions with Jeans length greater than the effective
driving wavelength of the turbulence. The transfer of power to
smaller scales might increase the ability of turbulence driven at
large scales to support even small-scale regions against collapse.
Computations including self-gravity that may confirm this are
described by Mac Low, Heitsch, \& Klessen (1999).

\subsection{The velocity space}

The $\Delta$-variance measuring the density structure of the
HD/MHD simulations can be compared directly to the analysis of 
astrophysical maps taken in optically thin tracers. However,
there is much additional information in the velocity space
which has to be addressed too.

Here, the $\Delta$-variance cannot be applied to the observations
since they retrieve only the line-of-sight integrated
one-dimensional velocity component convolved with the density.
Nevertheless, we can apply it to analyze the characteristic
quantities in the simulations where we have the full information
on the spatial distribution of the velocity vectors.
As the $\Delta$-variance measures the relative amount of structure
on certain scales in the density cubes it can be applied in
the same way to the velocity components or the energy density.

\begin{figure}
\epsfig{file=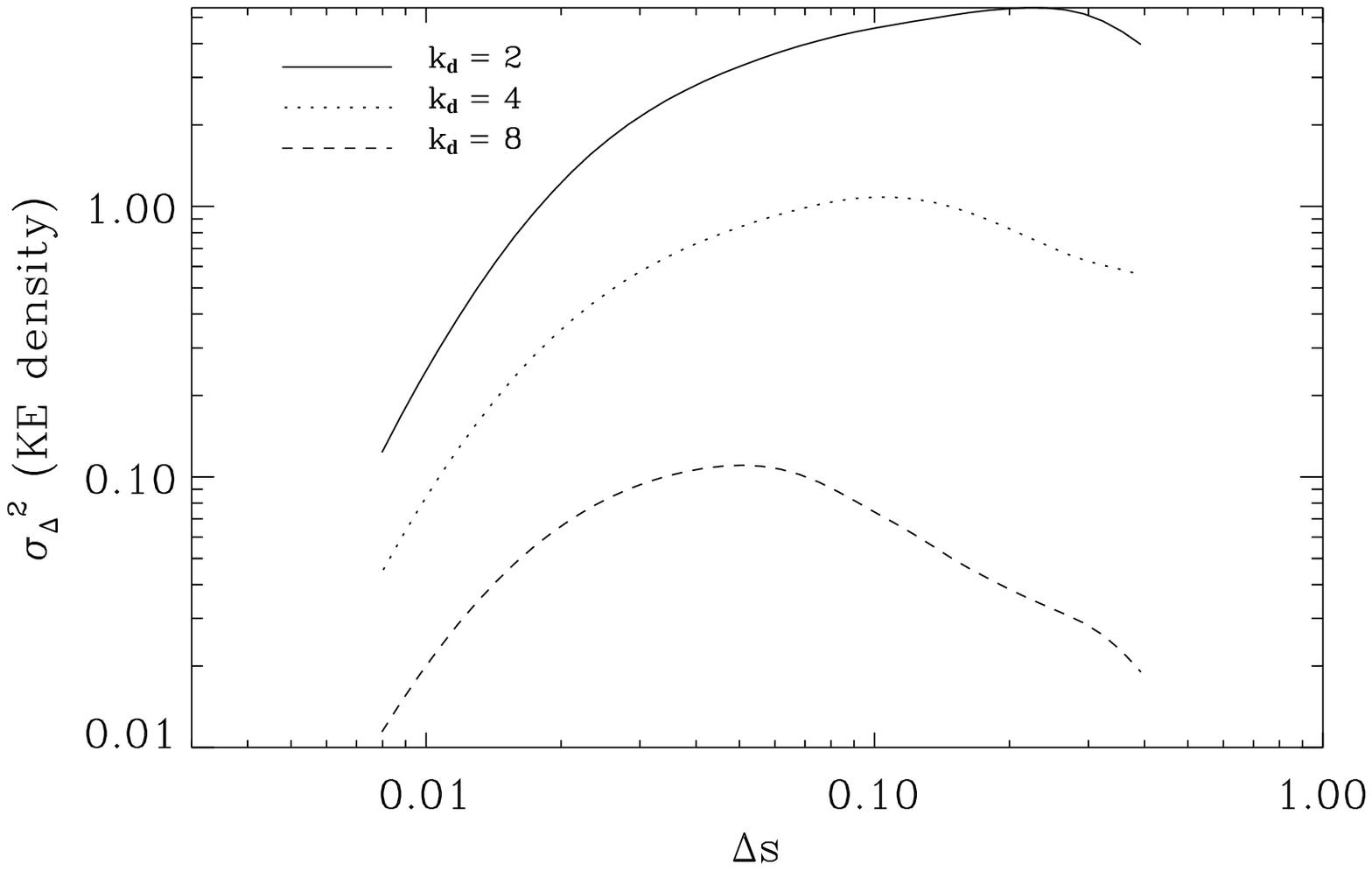,angle=0,width=\columnwidth}
\epsfig{file=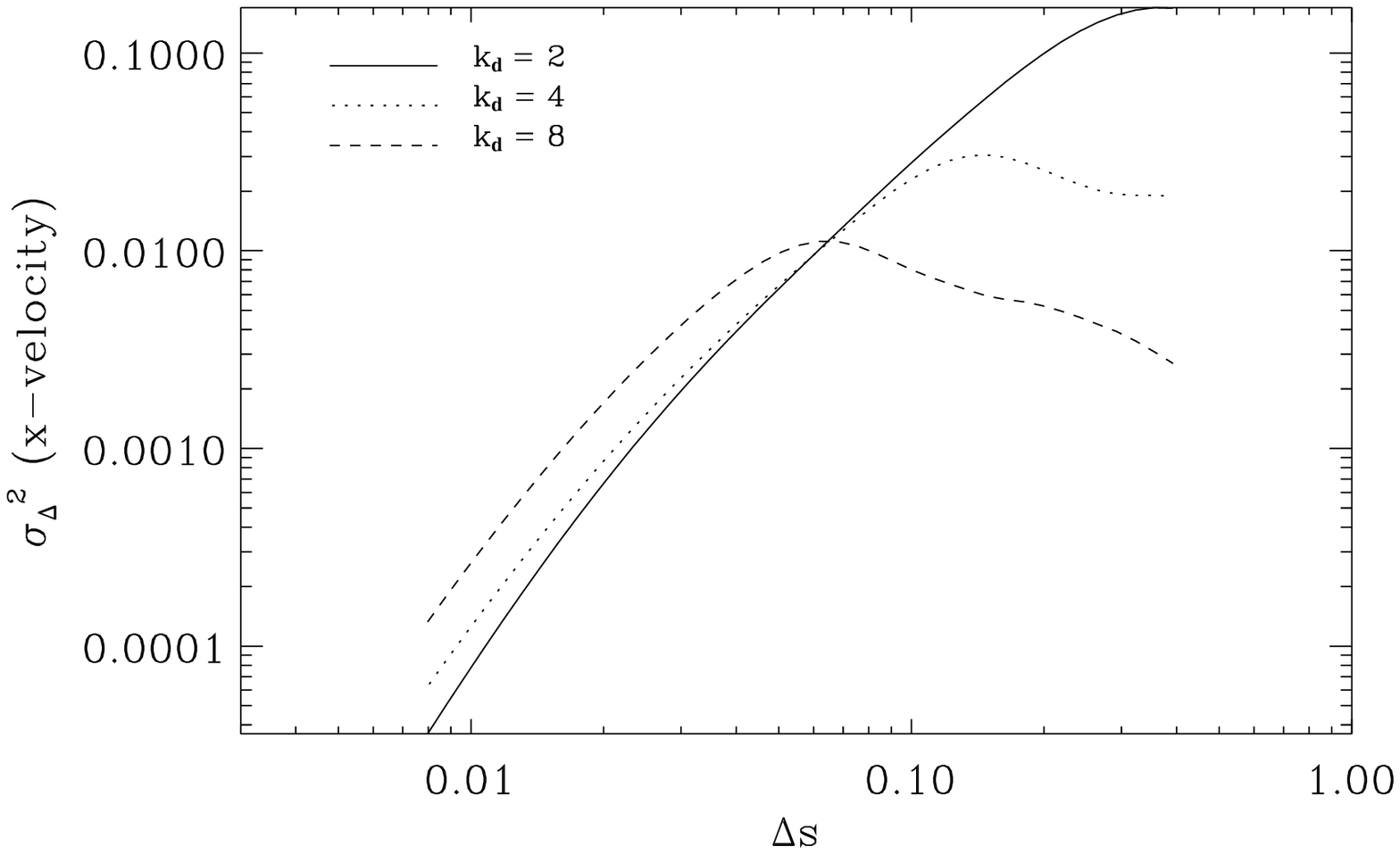,angle=0,width=\columnwidth}
\caption{$\Delta$-variances for the 
kinetic energy density, and the $x$-velocity component for the same models of
hydrodynamic turbulence driven with $k_d =2$, 4, and 8 shown in
Fig. \ref{driven_fig} (models HE2, HE4, and
HE8 from Mac Low 1999).
}
\label{energy_velocity}
\end{figure}

Fig. \ref{energy_velocity} shows the $\Delta$-variances for
for the kinetic energy density, and the $x$-velocity component of
the driven hydrodynamic model discussed in Sect. \ref{driven_sect}.
The plots can be compared to the $\Delta$-variances of the corresponding
density structures shown in the upper part of Fig. \ref{driven_fig}.
We see a shift of the dominant structure size from the driving wavelength
that is directly seen in the velocity structure to smaller scales
for the density structure. The energy density structure shows an
intermediate  behavior as a combination of density and velocity structure.

The same comparison for the supersonic model shown in the lower
part of Fig. \ref{driven_fig} provides much smaller differences
in the peak position of the $\Delta$-variance for the three
quantities.
This means that hypersonic velocities are not able to create
density structures at the scale of injection but only on some smaller
scales whereas smaller velocities produce void and compressed regions
directly at the scale of their occurrence.

The slopes in the self-similar range at smaller scales are
different for the density and velocity structure. The Gaussian 
perturbations in velocity space
create a $\Delta$-variance slope of 2.1 in the projected velocities
but do not translate into the same structural variations in 
the other quantities. In the density structure,
perturbations are created more efficiently at smaller scales
so that we obtain a slope of 0.45. The energy density
structure turns out to be dominated by the density variations
so that we find about the same slope there.

The difference in the $\Delta$-variances between the three
quantities is however probably due to the special driving mechanism.
If we apply the same analysis to the decaying turbulence models, we find 
that the peak position and slopes in all three quantities approach each 
other after some 
time, so that an equipartition of structure in density and velocity is
produced. In the first steps of the decaying model from Fig. 
\ref{decay_fig} we still find a difference in the slopes of the 
$\Delta$-variances between the density and velocity structure of a factor 
1.5 to 2 whereas the slopes are almost identical at the latest step.
Applying the same line of reasoning to the astrophysical 
observations, the comparison of density and velocity structure there might
help to clarify the state of relaxation and the driving mechanism
creating structure in interstellar clouds.

\section{Conclusions and Outlook}

\subsection{Conclusions}

In this paper we have shown that wavelet transform methods, as
exemplified by the $\Delta$-variance described by Stutzki et
al. (1998), offer a useful tool for comparison of observed structure
in molecular clouds to simulations of magnetized turbulence.  The
$\Delta$-variance spectrum can be analytically related to the more
commonly used Fourier power spectrum, but has distinct advantages: it
explicitly reveals finite map size and finite resolution effects; it
works in the absence of periodic boundary conditions; and it will
reveal characteristic structure scale even in the presence of shocks
and other sharp discontinuities.  One note of caution is called for in
its use, however: 2D spectra are proportional to the 3D spectra
multiplied by the lag, and this can introduce apparent power-law
behavior even in cases where the 3D spectra do not appear to have any
such behavior intrinsically.

We computed $\Delta$-variance spectra for the numerical
simulations of compressible, hydrodynamical and MHD turbulence
described by Mac Low et al. (1998) in the freely decaying case, and by
Mac Low (1999) in the case of driven turbulence, along with a few
extra models run to expand the parameter space in interesting
directions.  Resolution studies reveal that the $\Delta$-variance
spectra cleanly pick out the scale on which artificial viscosity
operates, which appears as a steeply dropping section of the spectrum
at small lags.  Examination of spectra from widely different times for driven
models in equilibrium shows that the $\Delta$-variance spectrum offers
a stable characterization of the dynamically varying structure.

Decaying hydrodynamical turbulence excited initially with a range of
length scales only appears to have self-similar, power-law behavior in
the hypersonic regime.  Once the rms Mach number drops below $M
\sim 4$ or so, a distinct length scale appears that grows as the
square root of time.  This appears to confirm the prediction made by Mac Low
(1999) that the effective driving scale must increase to explain the
inverse linear dependence of the kinetic energy dissipation rate on
the time.

Driven hydrodynamical turbulence can maintain self-similar, power-law
behavior at scales less than the driving scale, with a slope that lies
directly in the range of slopes observed for real molecular clouds.
In the observations, such power-laws extend to the largest scales in the
map that can be analyzed, suggesting that driving mechanisms may be
acting that add power on scales larger than those of the individual
clouds and clumps that are mapped.

Molecular clouds are observed to have magnetic fields strong enough
for the Alfv\'en velocities to be of the same order of magnitude as
the observed rms velocities (e.g. Crutcher 1999).  We have therefore
examined the effects of magnetic fields on our results from the
hydrodynamic models. 
We find that even strong magnetic fields often have fairly
small effects, but that they do tend to transfer power from large to
small scales, with implications for the support of small Jeans
unstable regions by large-scale driving mechanisms.
Contrary to some expectations, we find that
magnetic fields do not tend to create self-similar behavior, but
rather tend to destroy it, but that weaker fields appear to do so
less.  Hypersonic turbulence with Alfv\'en numbers of a few appears to
be consistent with the observations of both power-law behavior and
relatively strong magnetic fields.

The combined analysis of the velocity and density structure in molecular clouds
can help to distinguish between the possible mechanisms driving interstellar
turbulence and to provide information on the internal relaxation or
virialization of the clouds on different scales.

\subsection{Outlook}

The next step in this work is to move from a general characterization
of supersonic turbulence to attempts to fit observations of specific
real interstellar clouds, using what we have learned so far to guide
our search.  This should yield constraints on the effective Mach and
Alfv\'en numbers in these clouds, and begin to show whether
supersonic, super-Alfv\'enic turbulence can indeed give a good
description of the structure of molecular clouds.

To get a detailed comparison between observations and simulations we
have to solve the full radiative transfer problem relating the
simulated structure to maps in common lines such as the lower CO
transitions, which are often optically thick. Having the full
radiative transfer computations also allows the fit to include not
just the observed map scaling relations but also the peak intensities,
line ratios, and line shapes, placing significant additional
constraints on the models.

Furthermore the structure analysis must be extended beyond the
investigation of isotropic scaling behaviour. Appropriate measures for
anisotropy or filamentarity, and the relationship between the density
and the velocity structure have to be found.  Our first results
presented here have only scratched the surface of the possibilities
for systematic comparison between cloud observations and direct
turbulence simulations.

\begin{acknowledgements}
We thank F. Bensch, A. Burkert, and J. Stutzki for useful
      discussions.  V.O. acknowledges support by the Deut\-sche
      For\-schungs\-ge\-sell\-schaft through the grant SFB 301C.
      Computations were performed at the Rechenzentrum Garching of the
      Max-Planck-Gesellschaft. ZEUS was used by courtesy of the
      Laboratory for Computational Astrophysics at the NCSA. This
      research has made use of NASA's Astrophysics Data System
      Abstract Service.
\end{acknowledgements}

\end{document}